# A NOVEL ULTRASONIC DEVICE FOR MONITORING IMPLANT CONDITION

Amirhossein Yazdkhasti, Ph.D.
Center for Surgical Innovation and Engineering
Department of Orthopaedic Surgery
Cedars Sinai Health System
amirhossein.yazdkhasti@cshs.org

Sophie Lloyd, Ph.D Candidate
Department of Biomedical Engineering
Dartmouth Thayer School of Engineering
sophie.lloyd2017@gmail.com

Joseph H. Schwab, M.D, MSc.
Center for Surgical Innovation and Engineering
Department of Orthopaedic Surgery
Cedars Sinai Health System
Joseph.Schwab@cshs.org

Miao Yu, Ph.D
Department of Mechanical Engineering
University of Maryland
mmyu@umd.edu

Hamid Ghaednia, Ph.D.
Center for Surgical Innovation and Engineering
Department of Orthopaedic Surgery
Cedars Sinai Health System
Hamid.ghaednia@csmc.edu

April 30, 2022

## ABSTRACT

Every year more than 2.3 million joint replacement is performed worldwide. Around 10% of these replacements fail those results in revisions at a cost of $8 billion per year. In particular patients younger than 55 years of age face higher risks of failure due to greater demand on their joints. The long-term failure of joint replacement such as implant loosening significantly decreases the life expectancy of replacement. One of the main challenges in understanding and treatment of implant loosening is lack of a low-cost screening device that can detect or predict loosening at very early stages. In this work we are proposing a novel method of screening implant condition via ultrasonic signals. In this method we are applying ultrasonic signals to the joint via several piezoresistive discs while reading signals with several other piezoresistive sensors. We are introducing a new approach

in interpreting ultrasonic signals and we prove in a finite element environment that our method can be used to assess replacement condition. We show how our new concept can detect and distinguish between different implant fixation failure types sizes and even locate the position of the failure. We believe this work can be a foundation for development of a new generation of ultrasonic diagnosis wearable devices. *K*eywords Acoustic · ultrasound · Screening · Diagnosis · Joint replacement · Implant

## 1 Introduction

Each year over 2.3 million patients worldwide benefit from joint replacement procedures of the knee, hip, shoulder, ankle, and other extremities [1, 2, 3, 4]. Currently around 6-12% of all joint replacement procedures involve revisions having a total cost of approximately $8 billion per year in the US [5, 6]. Both the replacement and revision operations are estimated to significantly increase by 175% and 137%, respectively, by 2030 [7, 8, 9]. In particular, patients younger than 55 years of age face an elevated risk of revision due to the greater demands placed on their joints as well as the steady increase in risk of implant fixation failure with in vivo duration [10, 11]. Regardless of the origin of the failure i.e. biological or mechanical [12, 9, 7, 13], failure of the fixation in cemented and uncemented implants yields to mechanical discontinuity at the implant-bone, or implant-cement or cement-bone interface [14, 15, 16].

The key constraint in understanding fixation failure is the lack of early screening, diagnosis and predictive methods. As an example, in early 2017, Bonutti et al. [17] reported high rates of early tibial component loosening in patients with a new implant design (15 knees, less than two years post-surgery). The loosening was seen via radiographic images only in two of the 15 knees. However, intra-operatively all knees were found to have grossly loose implants requiring revision.

Thus, the challenge is that failure initiates well before indicators become visible for current imaging gold standards. This also matches the mechanical understanding of interfaces between materials [18, 19, 20]. The interface failure initially starts at very early stages as micro cracks inside the material and then reach the interface and cases failure.

There are several efforts to solve this challenge: artificial intelligence (AI), smart implants, electrical impedance tomography, and acoustic emission. With the recent advancements in AI there has been efforts to automate detection of implant loosening on X-Rays and reduce the missed diagnosis of loosening cases [21, 22, 23]. However, these models are dependent on quality of imaging modalities and more importantly they are diagnostic methods rather than early screening methods. The wireless smart implants are another approach to increase the quality of life of patients following arthroplasty [24]. However as of now there are no smart implant capable of screening fixation quality. Moreover, these techniques can only affect the future patients and not the current patients. In one of our group's works, we developed a novel piezoresistive bone cement to monitor fixation quality. We integrated the concepts of machine learning, smart materials, and electrical impedance tomography [25, 26, 27]. However, this approach is only applicable for cemented metal implants or polymeric implants.

Another approach for detecting failure is Acoustic Emission (AE) in which acoustic sensors are used to detect specific patterns from the vibration of the musculoskeletal system [28]. Acoustic emission is a non-invasive method developed by Joseph Kaiser [29] that uses piezoresistive sensors to detect vibrations in a system. These signals/vibrations are then interpreted to assess structure health. Acoustic emission has been shown to be useful in a few areas of orthopaedic such as detecting implant wear [30, 31, 32], monitoring fracture [33, 34], and osteoarthritis [35, 36, 37] There are two main challenges in interpreting the AE signals: 1) signal attenuation, 2) interpretation of signals. Because the signals are generated from the vibrations in the defected joint or bone, they have very low amplitude, attune very quickly and therefore, very hard to detect. Moreover, we cannot control these signals hence their interpretation becomes very difficult.

In this work we propose a new method, in which we are using piezoresistive acoustic sensors to both actuate and sense mechanical vibrations/waves. In this method an electrode cuff with several piezoresistive sensors is tightened around the target joint, Fig. 1. One or multiple piezoresistive sensors are then used to apply low-frequency ultrasonic signals with different frequencies and in different patterns on the skin, while the rest of the sensors are used to read the reflected and refracted signals. Because we are using relatively low ultrasonic frequencies, we eliminate the effect of attenuation. Next the signals are analyzed using analytical or machine learning solutions, Fig.1(b) and a report of the fixation condition will be sent to the clinician. Moreover, because the input signals are controlled, the output signals can be predicted and analyzed easier.

In this study we prove this concept in a simulated finite element environment. We modeled more than 88,000 different scenarios of fixation health and created a new technique in visualizing ultrasonic signals. We called this signature images and showed how they can help us identify how different defects manifest the ultrasonic signals. We then showed that this concept has the potential for creating novel, low-cost wearable devices for screening of replacement fixation following arthroplasty.

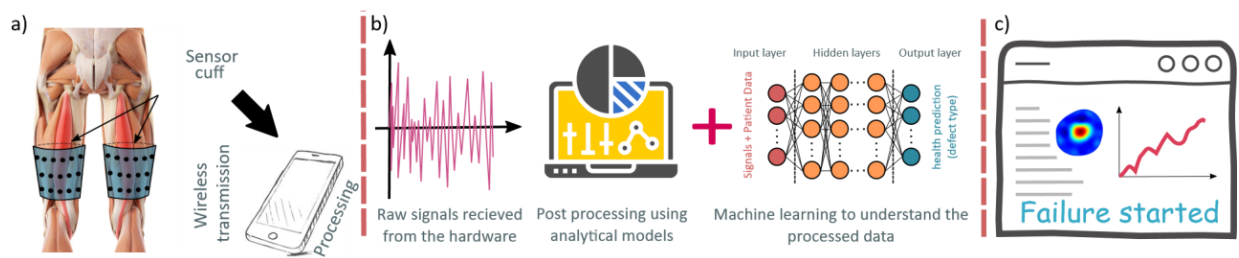

Figure 1: The concept. a) a wearable sensor cuffs is tightened around the target joint. The cuff is connected to a processing device i.e. smart phone or a computer. b) the signals are from the cuff are recorded and analyzed by the processing device. The analysis is a combination of theoretical and machine learning algorithms. The signals are interpreted to screen the replacement/implant condition. c) a report will be prepared for the clinician.

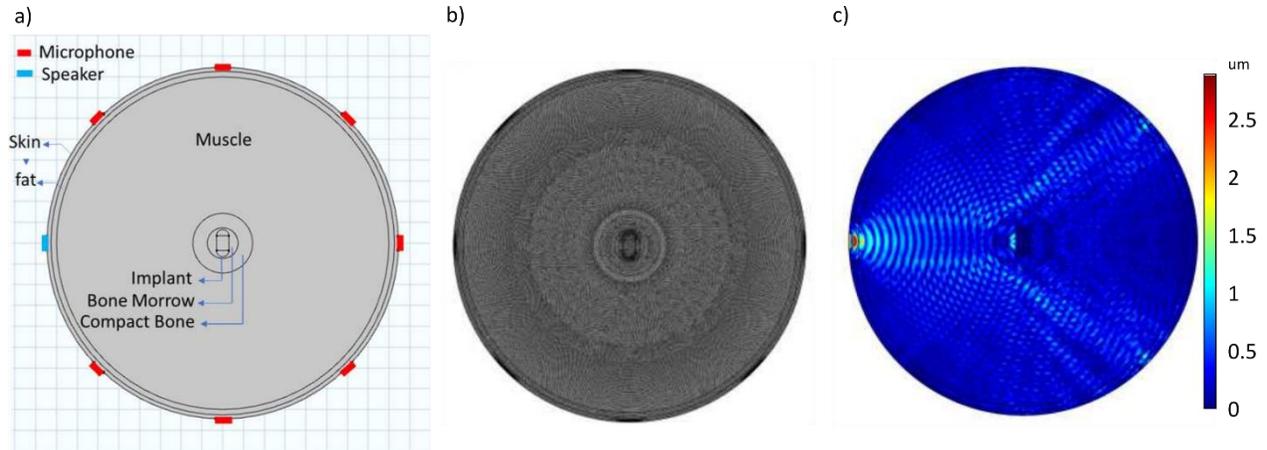

Figure 2: The modeling. a) the domain is modeled as a circle covered by skin and fat layers. the compact bone, bone morrow, and implant are placed at the center of the circle. b) the selected mesh following mesh convergence. The mesh includes 84314 domain elements and 3003 boundary elements. c) displacement field due to actuation from one of the sensors at 300 kHz.

| Material | Density, $\rho$ [kg/$m^3$] | Young's modulus [Pa] | Poisson's ratio, $v$ |
|---|---|---|---|
| Muscle | 1090 | 2.762e9 | 0.4 |
| Skin | 1109 | 2.900e9 | 0.29 |
| Fat | 911 | 1.889e9 | 0.29 |
| Compact bone | 1376 | 17e9 | 0.29 |
| Bone morrow | 115 | 0.520e9 | 0.29 |

Table 1: material properties used for different tissue types.

## 2 Methodology and Modeling

### 2.1 Concept and Modeling

In this stage of the study, we investigated monitoring fixation condition of a hip implant. The replacement was modeled in a 2-dimensional environment using COMSOL (COMSOL inc. MA, USA) Because addition of a 3rd dimension would not significantly enhance the study and would in reality decrease the accuracy due to limitation of element size in 3D models, we used a 2D environment. We modeled the human thigh as a circle of muscle covered by fat and skin layers. A ring of compact bone filled with bone morrow was then placed at the center of the circle/domain. Next the implant was placed at the center of bone morrow. All the nodes between the implant and bone morrow were completely connected for a perfect fixation. In this model we chose the material properties of different tissue types based on literature, Table 1. The implant was considered to be titanium.

Eight piezoresistive sensors are placed around the domain that can be used as both as actuators and sensors, Fig. 2(a). It has to be considered that these sensors are based on existing off-the-shelf piezoresistive disk sensors. The mesh convergence was performed at frequency of 300k and finally the mesh shown in Fig. 2(b) was selected. In total the selected mesh consists of 84314 domain elements and 3003 boundary elements with minimum element size of 3.3 m. Figure.2(c) shows an example of the displacement field due to actuation by one of the piezoresistive sensors.

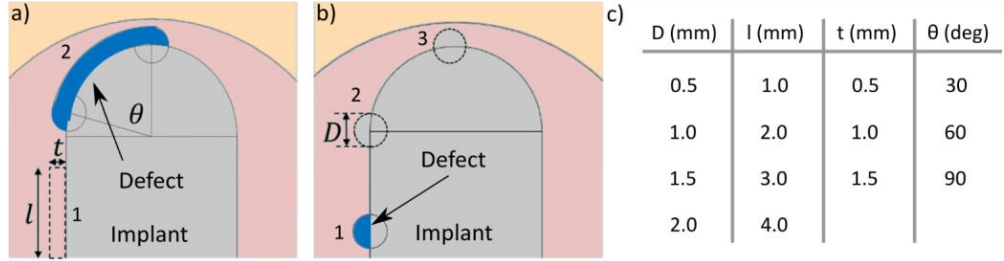

Figure 3: Defect types, their positions and parameters that were investigated. All possible combinations of positions, frequencies, geometrical parameters, and all possible actuation patterns of the sensors were simulated.

We investigated two types of defects, crack and loosening. The crack was modeled as a water-filled bubble at the interface of the implant and the bone morrow. The loosening was modeled as a thin layer of water at the interface of the implant and bone morrow. Several factors were investigated for both of these defect types. For loosening, Fig. 3(a) we investigated all possible combinations of two positions, three different thickness, and four different lengths, Fig. 3(c). For crack due to the radial symmetry of the model we investigated all possible combinations of three positions and four diameters, Fig. 3(b) and Fig. 3(c). For each condition we are also simulating actuation by each one of the eight piezoresistive sensors. For each case, we investigated 300 different frequencies from 1 kHz to 300 kHz with increments of 1kHz. At each frequency and condition, we are recording the signals received by all of the sensors that are not actuating the system. We are recording both amplitude and phase difference of the signals.

## 2.2 Data processing

In this method we perform a full rotation of actuating different sensors while recording signals from the other sensors. For example, for the model showed in Fig. 2(a) that we have eight sensors, we start by actuating one of the sensors and reading from the other seven. In the next cycle we actuate the sensor beside the first sensor and read from all other seven sensors. This goes on until all sensors actuated once. It is important to know that even though we analyze the whole displacement field as seen in Fig. 4(a) for all 300 different frequencies, in reality we only have access to the displacement at the sensors, Fig. 4(b). Figure 4(b) shows the data we capture from the sensors when one of the sensors is actuated. Even though the differences between healthy, cracked, and loose conditions are clear especially for frequencies higher than 200kHz, it is very challenging to find any pattern in the differences using these plots.

To overcome this challenge instead of using signals from the sensors individually, we created a matrix of data from all reading of all cycles of actuation. Therefore, for each set of simulation we record a matrix of 64x300 data points. Because we are looking at differences between different health conditions, we subtract the healthy data from each of the other conditions. This matrix can now be represented as an image, which we called signature image. Figure 5 shows an example of a signature image for a fixation with a crack compared to healthy fixation. Each row in the image includes 300 data points that are representing the amplitude of the signals at different frequencies. In each column 64 data point exist that includes readings from all eight sensors at each cycle of actuation. These images can be made for both amplitude and phase difference of the signals. In this study we are using these signature images to compare different fixation conditions. A perfect fixation would have a completely black image. Any noise or specific pattern on the signature images indicates a change in the fixation condition.

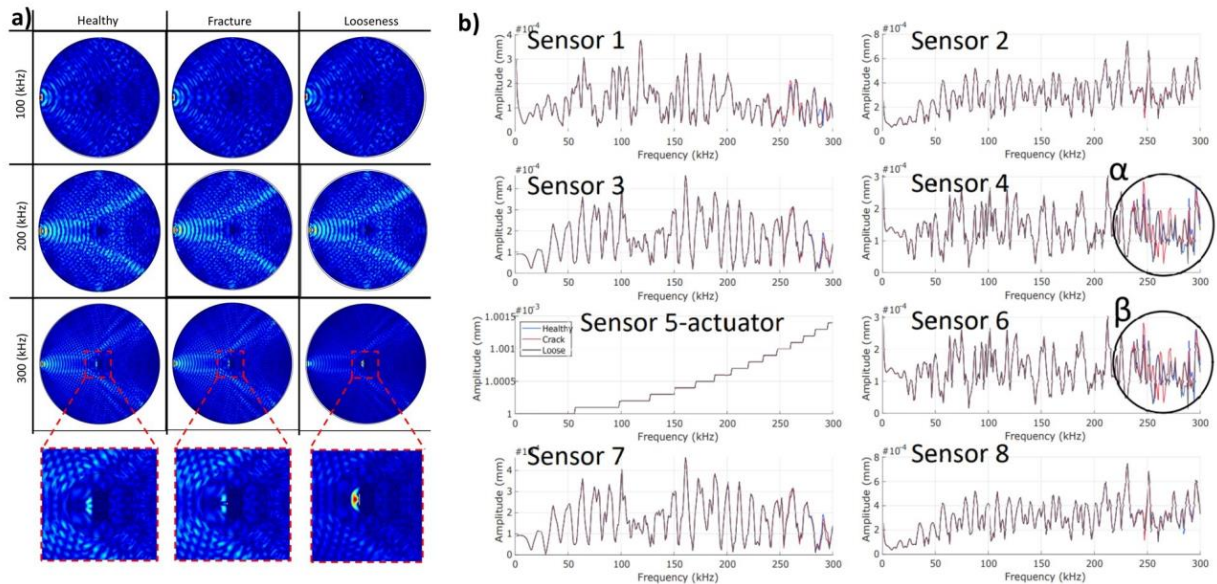

Figure 4: a) displacement field in the domain resulted from actuating the fifth sensor. As the frequency increases the wavelength decreases, therefore signals can detect finer details. b) the signals received by the sensors. It can be seen that at frequencies higher than 200 kHz the difference between the healthy and defected models becomes visible.

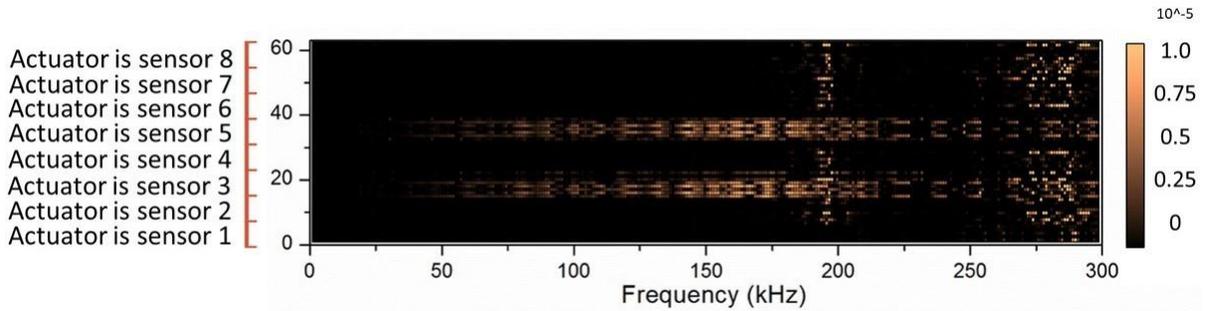

Figure 5: a) displacement field in the domain resulted from actuating the fifth sensor. As the frequency increases the wavelength decreases, therefore signals can detect finer details. b) the signals received by the sensors. It can be seen that at frequencies higher than 200 kHz the difference between the healthy and defected models becomes visible.

## 3 Results and Discussion

We investigated the signature images for different conditions to assess if our proposed concept is able to: 1) distinguish between different defect types, 2) assess defect size or severity, and 3) if signature images are able to locate the location of small defects. Figure 6 shows two example signature images of a cracked and a loose fixation. In Fig. 6(a)&(c) schematic of the defects and their location is showed. For a cracked (minor defect) fixation we have placed a water-filled for frequencies larger than 25 kHz several vertical patterns can be seen in the signals. bubble at the middle of the implant-bone interface. For the loose condition the loosening is placed at the tip of the implant and is modeled as a layer of water between the implant and bone.

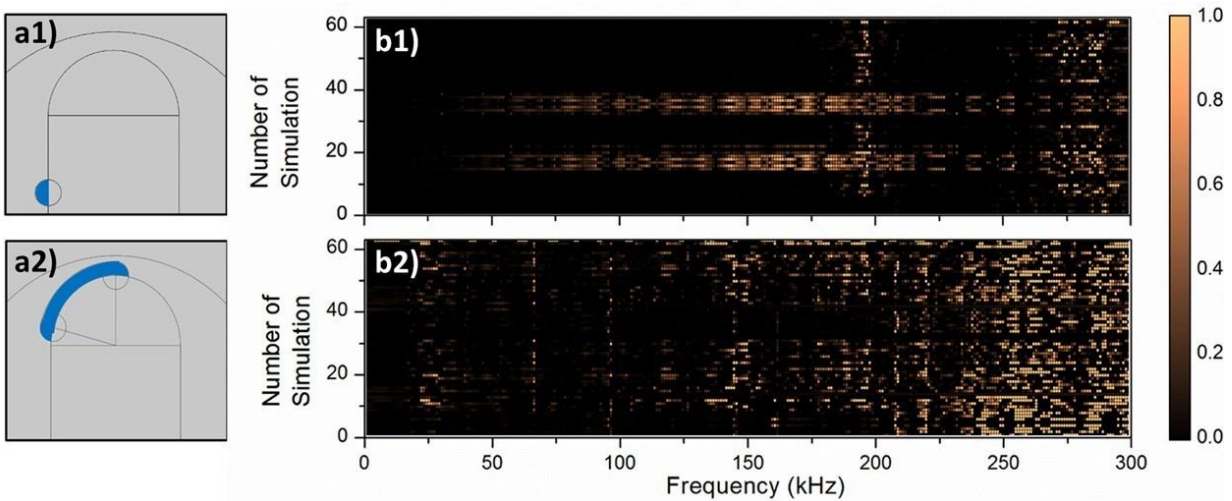

Figure 6: Signature images of a cracked fixation compared to a loose implant. a) a water-filled bubble is placed at the interface of the implant and bone morrow. b) signature image for the cracked fixation shown in Fig. 6(a). Two horizontal patterns can be seen for frequencies larger than 50 kHz. At frequencies larger than 200kHz vertical patterns start to appear. c) Schematic of a loose implant. d) signature image for a loose implant shown in Fig. 6(c).

Figure 6(b) shows the signature image for the cracked fixation. Two horizontal patterns can be seen in the signature image. We later show that the horizontal patterns are indications of the defect location. These patterns can be seen from frequencies larger than 25 kHz. This is very interesting because the wavelength of acoustic waves at small frequencies is relatively large and therefore, they cannot be used for imaging. However, here we show that even at small frequencies the ultrasound waves carry information of very fine details in their path. At frequencies larger than 200 kHz vertical patterns become visible which indicates all sensors are seeing the defects. Figure 6(d) is an example of a signature image for a loose implant shown in Fig. 6(c). The horizontal patterns cannot be seen anymore, however the defects are affecting the signal readings of all sensors from even frequencies as small as 25 kHz. The difference between the two health conditions is very clear.

Figure 7 is an example of the effect of defect size on signature images of a cracked implant. Figures 7(a1 to a4) are the schematics of the models, Figs. 7(b1 to b4) and (c1 to c4) are showing the amplitude and phase difference signature images respectively. From a1 to a4 the size of the defect increases gradually from 0.5 mm to 2.0 mm. The effect of size on the amplitude signature images is clear, as the defect size increases the amount of disturbance in the signature image increase. Moreover, as the defect size increases the changes in the signature image can be seen from lower frequencies. Interestingly the horizontal patterns show no change as the defect size increases. The phase difference signature images seem to be independent of the defect size.

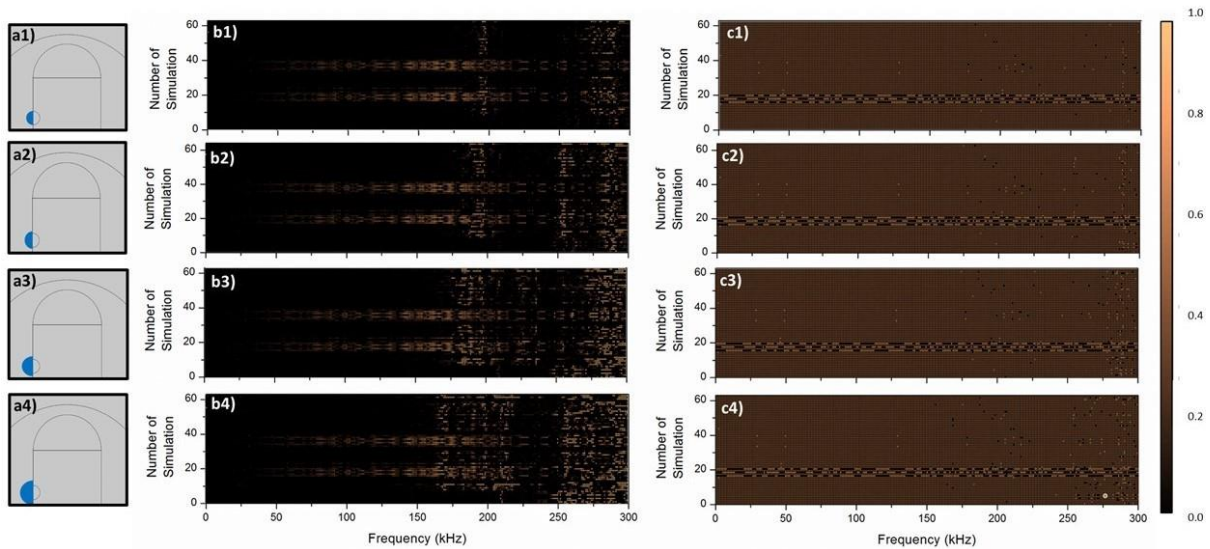

Figure 7: Effect of defect size on signature images. a1-4) schematics of the defects. b1-4) amplitude signature images of the cracked fixations showed in (a1-4). c1-4) phase difference signature images of (b1-4).

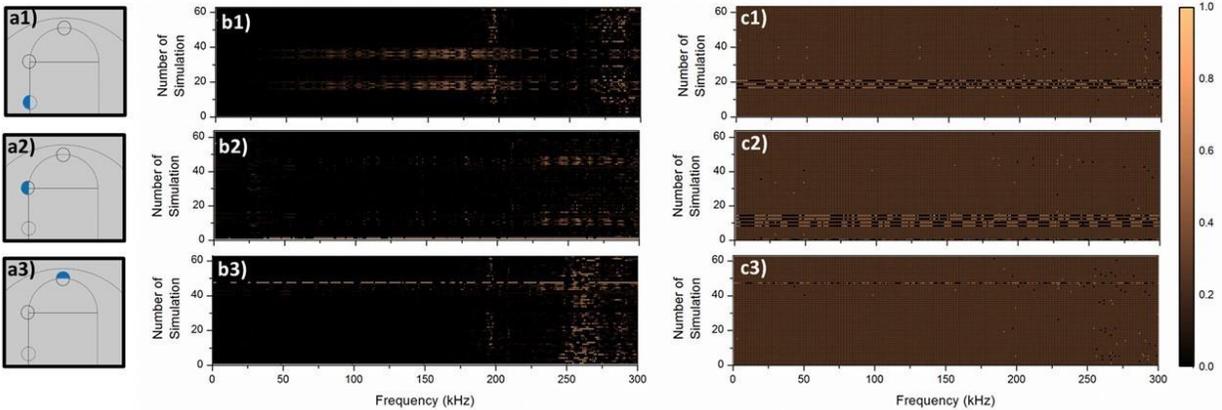

Figure 8: Effect of defect location on signature images. a1-3) schematics of the defects. b1-3) amplitude signature images of the cracked fixations showed in (a1-3). c1-3) phase difference signature images of (b1-3).

In Fig. 8 we are investigating the effect of defect location on the signature images. Figures 7(a1 to a3) are the schematics of the models, Figs. 8(b1 to b3) and (c1 to c3) are showing the amplitude and phase difference signature images respectively. In contradiction to Fig. 7 the horizontal patterns in the amplitude signature images change as the location of the defect change. As the defect moves from the side of the implant to the tip of the implant the distance between the two horizontal patterns increases. When the defect is at the top of the implant, there is only one horizontal pattern. The disturbance in the amplitude signature images do not show significant changes, which is an indication of them depending on the defect size rather than defect location. The phase difference signature images, Fig. 8(c1 to c-3), show very large changes as the defect moves. This shows that the patterns seen in phase difference images are dependent of the location.

## 4  Conclusion and Future work

In this work we investigated a novel method of assessing implant fixation using low-frequency low-intensity ultrasonic waves. We are using piezoresistive discs to generate and sense acoustic waves. These sensors could be assembled as a wearable cuff that can get connected to any computing device such as smart phones to monitor and screen fixation condition.

We developed a new way of representing ultrasonic signals, signature images, and we showed without the need of using very high frequencies ultrasonic signals are able to assess implant fixation. In this work we showed in a simulation environment that with our proposed method we can distinguish between different defect types, assess severity of defects and even locate the defects. Even though we showed that signature images reveal significant information about the defects and fixation condition of implants, there is still a lot we don't understand about this system and several aspects that need improvements. In future works we are planning to improve upon two aspects of this work, 1) modeling geometries closer to human body, and 2) validation on human cadavers.

References

[1] 2013 extremity update. *Orthop Newt News*, 24(1):1–16, 2013.

[2] 2013 hip and knee implant review. *Orthop Newt News*, 24(3):1–20, 2013.

[3] Steven M Kurtz, Kevin L Ong, Edmund Lau, Marcel Widmer, Milka Maravic, Enrique Gómez-Barrena, Maria de Fátima de Pina, Valerio Manno, Marina Torre, William L Walter, et al. International survey of primary and revision total knee replacement. *International orthopaedics*, 35(12):1783–1789, 2011.

[4] Orthoworld. *The Orthopedic Industry Annual Report*, 2015.

[5] Mohit Bhandari, Jon Smith, Larry E Miller, and Jon E Block. Clinical and economic burden of revision knee arthroplasty. *Clinical Medicine Insights: Arthritis and Musculoskeletal Disorders*, 5:CMAMD–S10859, 2012.

[6] Steven M Kurtz, Kevin L Ong, Jordana Schmier, Fionna Mowat, Khaled Saleh, Eva Dybvik, Johan Kärrholm, Göran Garellick, Leif I Havelin, Ove Furnes, et al. Future clinical and economic impact of revision total hip and knee arthroplasty. *JBJS*, 89(suppl_3):144–151, 2007.

[7] Michael Pitta, Christina I Esposito, Zhichang Li, Yuo-yu Lee, Timothy M Wright, and Douglas E Padgett. Failure after modern total knee arthroplasty: a prospective study of 18,065 knees. *The Journal of arthroplasty*, 33(2):407–414, 2018.

[8] Steven M Kurtz, Kevin L Ong, Edmund Lau, and Kevin J Bozic. Impact of the economic downturn on total joint replacement demand in the united states: updated projections to 2021. *JBJS*, 96(8):624–630, 2014.

[9] Connor Kenney, Steven Dick, Justin Lea, Jiayong Liu, and Nabil A Ebraheim. A systematic review of the causes of failure of revision total hip arthroplasty. *Journal of orthopaedics*, 2019.

[10] S Adelaide. *Australian Orthopaedic Association National Joint Replacement Registry, 2016*.

[11] H Hempstead. *National Joint Registry for England W, Northern Ireland and the Isle of Man, 14th Annual Report.; 2017.*

[12] Joshua J Jacobs, Kenneth A Roebuck, Michael Archibeck, Nadim J Hallab, and Tibor T Glant. Osteolysis: basic science. *Clinical Orthopaedics and Related Research®*, 393:71–77, 2001.

[13] Anne Postler, Cornelia Lützner, Franziska Beyer, Eric Tille, and Jörg Lützner. Analysis of total knee arthroplasty revision causes. *BMC musculoskeletal disorders*, 19(1):55, 2018.

[14] No authors listed. 14th annual report, 2017. national joint registry for england, wales, northern ireland and the isle of man (njr). 2017.

[15] Denis Nam, Charles M Lawrie, Rondek Salih, Cindy R Nahhas, Robert L Barrack, and Ryan M Nunley. Cemented versus cementless total knee arthroplasty of the same modern design: a prospective, randomized trial. *The Journal of bone and joint surgery. American volume*, 101(13):1185, 2019.

[16] CM Lawrie, M Schwabe, A Pierce, RM Nunley, and RL Barrack. The cost of implanting a cemented versus cementless total knee arthroplasty. *The Bone & Joint Journal*, 101(7_Supple_C):61–63, 2019.

[17] Peter M Bonutti, Anton Khlopas, Morad Chughtai, Connor Cole, Chukwuweike U Gwam, Steven F Harwin, Brent Whited, Didi E Omiyi, and Joshua E Drumm. Unusually high rate of early failure of tibial component in attune total knee arthroplasty system at implant–cement interface. *The journal of knee surgery*, 30(05):435–439, 2017.

[18] Hamid Ghaednia, Sara A Pope, Robert L Jackson, and Dan B Marghitu. A comprehensive study of the elastoplastic contact of a sphere and a flat. *Tribology International*, 93:78–90, 2016.

[19] Hamid Ghaednia, Xianzhang Wang, Swarna Saha, Yang Xu, Aman Sharma, and Robert L Jackson. A review of elastic-plastic contact mechanics. *Applied Mechanics Reviews*, 2017.

[20] Hamid Ghaednia, Dan B Marghitu, and Robert L Jackson. Predicting the permanent deformation after the impact of a rod with a flat surface. *Journal of Tribology*, 137(1):011403, 2015.

[21] Romil F Shah, Stefano A Bini, Alejandro M Martinez, Valentina Pedoia, and Thomas P Vail. Incremental inputs improve the automated detection of implant loosening using machine-learning algorithms. *The Bone & Joint Journal*, 102(6 Supple A):101–106, 2020.

[22] Alireza Borjali, Antonia F Chen, Orhun K Muratoglu, Mohammad A Morid, and Kartik M Varadarajan. Detecting total hip replacement prosthesis design on preoperative radiographs using deep convolutional neural network. *arXiv preprint arXiv:1911.12387*, 2019.

[23] Christophe Nich, Julien Behr, Vincent Crenn, Nicolas Normand, Harold Mouchère, and Gaspard d'Assignies. Applications of artificial intelligence and machine learning for the hip and knee surgeon: current state and implications for the future. *International Orthopaedics*, pages 1–8, 2022.

[24] Cody O'Connor and Asimina Kiourti. Wireless sensors for smart orthopedic implants. *Journal of Bio-and Tribo-Corrosion*, 3(2):1–8, 2017.

[25] Hamid Ghaednia, Crystal E Owens, Tyler N Tallman, Anastasios J Hart, and Kartik M Varadarajan. Non-invasive diagnosis of aseptic implant loosening via electrical impedance tomography. *International Society of Technology in Arthroplasty (Toronto, CN)*, 2019.

[26] Hamid Ghaednia, Crystal Owens, Ricardo Roberts, Tyler N Tallman, Anastasios John Hart, and Kartik Mangudi Varadarajan. Interfacial load monitoring and failure detection in total joint replacements via piezoresistive bone cement and electrical impedance tomography. *Smart Materials and Structures*, 29:085039, 2020.

[27] Hamid Ghaednia, Crystal E Owens, Lily E Keiderling, Kartik M Varadarajan, A John Hart, Joseph H Schwab, and Tyler T Tallman. Is machine learning able to detect and classify failure in piezoresistive bone cement based on electrical signals? *arXiv preprint arXiv:2010.12147*, 2020.

[28] KA Olorunlambe, DET Shepherd, and KD Dearn. A review of acoustic emission as a biotribological diagnostic tool. *Tribology-Materials, Surfaces & Interfaces*, 13(3):161–171, 2019.

[29] Hans Maria Tensi. The kaiser-effect and its scientific background. *Journal of Acoustic Emission*, 22:s1–s16, 2004.

[30] Hans-Joachim Schwalbe, Guido Bamfaste, and RP Franke. Non-destructive and non-invasive observation of friction and wear of human joints and of fracture initiation by acoustic emission. *Proceedings of the Institution of Mechanical Engineers, Part H: Journal of Engineering in Medicine*, 213(1):41–48, 1999.

[31] RP Franke, P Dorner, HJ Schwalbe, and B Ziegler. Acoustic emission measurement system for the orthopedic diagnostics of the human femur and knee joint. *Journal of Acoustic Emission*, 22:236–242, 2004.

[32] Ataif Khan-Edmundson, Geoffrey W Rodgers, Tim BF Woodfield, Gary J Hooper, and J Geoffrey Chase. Tissue attenuation characteristics of acoustic emission signals for wear and degradation of total hip arthroplasty implants. *IFAC Proceedings Volumes*, 45(18):355–360, 2012.

[33] Serife Agcaoglu and Ozan Akkus. Acoustic emission based monitoring of the microdamage evolution during fatigue of human cortical bone. *Journal of biomechanical engineering*, 135(8), 2013.

[34] C Van Toen, J Street, Thomas R Oxland, and Peter Alec Cripton. Acoustic emission signals can discriminate between compressive bone fractures and tensile ligament injuries in the spine during dynamic loading. *Journal of biomechanics*, 45(9):1643–1649, 2012.

[35] L-K Shark, H Chen, and John Goodacre. Knee acoustic emission: a potential biomarker for quantitative assessment of joint ageing and degeneration. *Medical engineering & physics*, 33(5):534–545, 2011.

[36] Tawhidul Islam Khan and Harino Yoho. Integrity analysis of knee joint by acoustic emission technique. *Journal on Multimodal User Interfaces*, 10(4):319–324, 2016.

[37] Lik-Kwan Shark, Hongzhi Chen, and John Goodacre. Discovering differences in acoustic emission between healthy and osteoarthritic knees using a four-phase model of sit-stand-sit movements. *The open medical informatics journal*, 4:116, 2010.